# Towards quantifying the role of exact exchange in predictions of transition metal complex properties

Efthymios I. Ioannidis[1] and Heather J. Kulik[1,*]

[1]Department of Chemical Engineering, Massachusetts Institute of Technology, Cambridge, Massachusetts, 02139 United States

We estimate the prediction sensitivity with respect to Hartree-Fock exchange in approximate density functionals for representative Fe(II) and Fe(III) octahedral complexes. Based on the observation that the range of parameters spanned by the most widely-employed functionals is relatively narrow, we compute electronic structure property and spin-state orderings across a relatively broad range of Hartree-Fock exchange (0-50%) ratios. For the entire range considered, we consistently observe linear relationships between spin-state ordering that differ only based on the element of the direct ligand and thus may be broadly employed as measures of functional sensitivity in predictions of organometallic compounds. The role Hartree-Fock exchange in hybrid functionals is often assumed to play is to correct self-interaction error-driven electron delocalization (e.g. from transition metal centers to neighboring ligands). Surprisingly, we instead observe that increasing Hartree-Fock exchange reduces charge on iron centers, corresponding to effective delocalization of charge to ligands, thus challenging notions of the role of Hartree-Fock exchange in shifting predictions of spin-state ordering.

## I. Introduction

Density functional theory (DFT) has seen widespread use and exponential growth[1] owing to its relatively computationally efficient description of short-range, dynamic correlation. Ease of entry for new users has made practical DFT one of the most popular "black box" computational chemistry techniques, despite well-known shortcomings. Namely, predictions are sensitive to user selection of the exchange-correlation functional amongst a "zoo" of choices. Decisions about functional choice are in turn often influenced by word of mouth or popular opinion[2] and availability in a localized basis or plane wave electronic structure code. Extensive optimization of exchange-correlation functional parameters in DFT against test sets with a large number of parameters[3] has improved accuracy, though reduction in parameters[4] or limited use of parameters in some functionals[5,6] can provide improved transparency. Nevertheless, mathematical expressions for exchange and correlation still prevent a clear understanding of how accuracy may be systematically and globally improved. Established test and training sets for functional development primarily focus on thermochemistry of main group molecules[7], and accuracy is not necessarily transferable to other properties or elements.

Importantly, exchange-correlation functionals that work well for main group elements may not work as well for transition metals[8], which are central to homogeneous[9-11], heterogeneous[12], or enzymatic[13] catalysis. Transition metals are increasingly prominent in a computational design screens[12,14] for which high-accuracy and high-efficiency "black box" DFT predictions are needed. Nevertheless, transition metals remain a challenge due to the close spacing of electron configurations (e.g. $3d^74s^1$ vs. $3d^64s^2$ in neutral Fe) that leads to several accessible spin states and oxidation states[15]. Spin-crossover (SCO) complexes[16,17], which typically contain Fe(II) or Fe(III) centers[18], represent a particularly challenging class of molecules because the spin state can change with small changes in temperature. SCO molecules have shown promise in nanoscale storage devices[19], spintronics[20,21], and catalysis[22-24]. Nevertheless, common exchange-correlation functionals struggle to reproduce critical features for the spin-dependent potential energy surfaces[25-28].

In SCO molecules, low-spin states are known to be favored by generalized gradient approximation exchange-correlation functionals, while hybrid functionals that include a fraction of Hartree-Fock exchange often prefer high-spin states[28-33], and different energy gaps are obtained with different exchange-correlation functionals[34,35]. One suggested reason for the failure of practical DFT in describing organometallics is that relatively localized 3d valence electrons suffer strongly from self-interaction error (SIE) present in pure DFT functionals and only approximately corrected in hybrid functionals. Strides have been made in systematic removal of SIE[36, 37] and identification of paths to improve balance in spin-state ordering[28, 29, 38-42], but hybrid functionals remain a popular and straightforward approach to approximately correct for energetic errors driven by imbalances in SIE between spin states.

It is worthwhile to note that mixing in of Hartree-Fock exchange may potentially trade reduction in self-interaction errors for an increase in static correlation errors, which tend to plague Hartree-Fock more than density functional theory approaches. For a balanced treatment of both static correlation and in the absence of self-interaction error, multireference wavefunction techniques have been used[43-50] to study spin crossover complexes up to around 45 atoms in size[48]. The predominant method employed in the study of spin-crossover complexes is CASPT2, and, while it scales more expensively than density functional theory approaches, recent improvements in scaling[44] have made larger systems tractable. Other studies have applied the even more expensively-scaling CCSD(T) to smaller spin crossover complexes[49]. Wavefunction approaches are not without challenges and in some cases still produce sizeable, 5 kcal/mol energetic errors in spin-state ordering[48]. However, they are typically in very good agreement with experimental spin crossover properties and are a suitable reference for benchmarking of exchange-correlation functionals for higher throughput studies. Despite advances in wavefunction theory, approximate density functionals are still preferred by most computational researchers due to ease of use and lower scaling that makes geometry optimization and high-throughput calculations feasible.

Extending study[31-33] of how organometallic complex spin states vary with functional choice can broaden an understanding of the ways in which hybrid functionals improve predictions of spin-state orderings, especially since both low[32, 51, 52] and high[31, 53, 54] percentages of Hartree-Fock exchange have been proposed for the description of transition metal complexes. Rather than focusing on finding one prescription for exchange, we aim to understand the way in which relative energetic, electronic, and structural properties of spin states are sensitive to these descriptions. Understanding this variability unifies many functionals and can provide a useful guide for interpreting the prediction bias introduced through functional choice in DFT literature. Finally, we aim to enlarge a quantitative understanding of how self-interaction error manifests and is balanced through the use of Hartree-Fock exchange.

The outline of the paper is as follows. In Sec. II, we discuss common descriptions of quantum mechanical exchange. Computational details are presented in Sec. III. Section IV contains preliminary observations on qualitative spin-state ordering for classes of exchange correlation functionals. Sec. V considers in detail correlations and trends of energetic, electronic, and geometric properties of organometallics under varied Hartree-Fock exchange, and we provide conclusions in Sec. VI.

## II. Descriptions of exchange

Exchange energies are a critical component in the prediction of relative spin-state ordering. Hartree-Fock (HF) theory provides exact treatment of Coulomb repulsion and exchange for a wavefunction described by a Slater determinant of single particle orbitals. The form of Hartree-Fock exchange energy is as follows:

$$E_x^{HF} = -\frac{1}{2} \sum_{i,j}^{occ} \int d\mathbf{r} d\mathbf{r}' \frac{\phi_i^*(\mathbf{r})\phi_j^*(\mathbf{r}')\phi_j(\mathbf{r})\phi_i(\mathbf{r}')}{|\mathbf{r}-\mathbf{r}'|} \qquad (1)$$

where $i$ and $j$ are same-spin, occupied molecular orbitals ($\phi$). For occupation of single particle orbitals at equivalent energies in closed shell (e.g. S=0) and open shell (e.g. S=1) complexes, the $E_x^{HF}$ term will lower the relative energy of the high-spin state. In most cases, there is a difference in single-particle orbital energies populated in low-spin and high-spin, which will counterbalance the exchange-driven stabilization. In mid-row transition metal complexes with weak ligand field splitting, the spread in orbital energies will be small, and the high-spin state will become the ground state. The exact treatment of exchange in Hartree-Fock theory is unfortunately paired with an absence of direct treatment of short-range, dynamic correlation, which is critical for bonding and barrier height estimation.

In contrast with Hartree-Fock, density functional theory (DFT) provides explicit inclusion of short-range correlation. However, this treatment of electron correlation in practice comes with a critical downside: the Coulomb repulsion energy is estimated as an integral of each single particle orbital with the density of the rest of the system, resulting in what is known as self-interaction error (SIE). Exchange in DFT must then do double duty by both reducing repulsion of same-spin electrons (as in HF) as well as eliminating SIE. The simplest approximation to exchange in DFT is the local-density approximation (LDA), which is given by:

$$E_x^{LDA}[\rho] = -\frac{3}{2}\left(\frac{3}{4\pi}\right)^{1/3} \sum_\sigma \int \rho_\sigma(\mathbf{r})^{4/3} d\mathbf{r} \quad . \tag{2}$$

LDA exchange is obtained from the assumption that the density maybe modeled locally as a homogeneous electron gas with the same density as the real system. For quickly varying densities, as in molecules with localized subshells, LDA exchange provides a particularly poor estimate of the exchange energy.

Beyond the LDA, gradients of the density may be directly incorporated into semi-local descriptions of exchange, typically rescaled by the absolute value of the density as in the so-called generalized gradient approximation (GGA). The B88[55] GGA exchange energy is given by:

$$E_x^{GGA} = E_x^{LDA} - \beta \sum_\sigma \int \rho_\sigma^{4/3} \frac{x_\sigma^2}{\left(1 + 6\beta x_\sigma \sinh^{-1} x_\sigma\right)} d\mathbf{r} \quad , \tag{3}$$

where this exchange energy is referenced with respect to the LDA energy and is a rescaled integral of the spin density (spin index σ) with a semi-empirical parameter $\beta = 0.0042$ a.u.. Here, the variable $x_\sigma$ is the rescaled gradient of the density:

$$x_\sigma = \frac{|\nabla \rho_\sigma|}{\rho_\sigma^{4/3}} \quad . \tag{4}$$

Neither LDA or GGA exchange are well-suited for correcting imbalances between spin states in the extent of excess Coulomb repulsion from SIE. That is, low-spin and high-spin states have differing fractions of delocalized (bonding) versus localized (non-or anti- bonding) states and therefore high-spin states are expected to be destabilized within LDA or GGA descriptions of exchange in DFT. Ganzenmüller and coworkers have verified this qualitative statement for a simple model system, $FeH_6$, by comparing the role of enhancing differing types of exchange on the spin-state splitting.[32] Instead, self-interaction correction schemes[36, 37] or delocalization tuning on a subshell with DFT+U[42] are pure DFT-based approaches for correcting SIE imbalances.

Hybrid functionals are a widely employed approach for approximately correcting self-interaction errors in practical DFT, largely made successful by effective error cancellation between the lack of electron correlation in HF with the SIE effects in practical DFT. One of the most well-known hybrid functionals, B3LYP[56-58], is defined as:

$$E_{xc}^{B3LYP} = E_x^{LDA} + a_0(E_x^{HF} - E_x^{LDA}) + a_x(E_x^{GGA} - E_x^{LDA}) + E_c^{LDA} + a_c(E_c^{GGA} - E_c^{LDA}) \quad (5)$$

where $a_0$=0.20, which corresponds to 20% Hartree-Fock exchange, and the GGA (B88) enhancement factors over LDA are $a_x$=0.72 and $a_c$=0.81 for exchange and correlation, respectively. Here, an admixture of Hartree-Fock (HF) exchange on the non-interacting, single particle orbitals in a DFT calculation is mixed with GGA and LDA descriptions of exchange. More recently, range-separated hybrids[59] have added additional tunable parameters that aid in distinct treatment of long-range exchange versus short-range exchange, but the central focus of this work is on short-range exchange on transition metal centers.

While B3LYP is commonly employed to successfully describe organic systems, its direct application to organometallics leads to mixed results. One approach is to adjust the extent of Hartree-Fock exchange in a functional in order to reproduce key energetics and spin-state orderings in organometallic systems where multiple spin multiplicities lie close in energy.[27, 31, 51-54] However, such exchange-correlation functional tuning is then constrained by the availability of experimental data or well-converged correlated quantum chemistry results. Importantly, the outcome from these fitting studies are often contradictory: alternative mixings of 0%[32], 15%[51, 52], 25%[53, 54], and 30-50%[31] HF exchange have all been proposed for Fe(II) octahedral complexes alone. Such broad outcomes suggest that the mixing of exact exchange in a functional is highly dependent on the underlying chemistry of the system, and a one-size-fits-all approach is not likely to be successful. Preliminary success has been made in identifying chemically-motivated ways to tune functional parameters[60-62] outside of organometallic chemistry. The appropriate tuning for organometallic complexes or correlated materials is largely still approximated on local measures of the chemical potential of the subshell of interest[63], and efforts to improve functionals on transition metal test sets have indicated no clear path to optimization[64].

**III. Computational Details**

Calculations were carried out using the TeraChem[65] package for all LDA, GGA, and GGA hybrid calculations. The default B3LYP definition in TeraChem uses the VWN1-RPA form for the LDA VWN[66] component of LYP correlation[56]. Initial calculations on GGA hybrids also considered the effect of using instead the 3-parameter or 5-parameter forms of the VWN correlation[66] (B3LYP3, B3LYP5 keywords) as well as using other forms of the correlation with the B3P86[57,67], B3PW91[57,68], PBE0[5,6] (25% HF exchange vs. 20% in B3LYP), or B97[69] (19% HF exchange) GGA hybrids. Overall qualitative GGA hybrid predictions were unchanged and therefore B3LYP1 is chosen as the representative functional.

Altered Hartree-Fock exchange percentages in a modified form of B3LYP were implemented in TeraChem for this work. Meta-GGA calculations were carried out with Q-Chem 4.2. All calculations were performed using the LANL2DZ effective core potential basis for the iron atom and the 6-31G* basis for the other atoms. Geometry optimizations were carried out using the L-BFGS algorithm in Cartesian coordinates, as implemented in DL-FIND[70], to default thresholds of $4.5 \times 10^{-4}$ hartree/bohr for the maximum gradient and $1 \times 10^{-6}$ hartree for the change in SCF energy between steps.

High-spin states (quintet multiplicity for Fe(II) and sextet for Fe(III)) are compared against low-spin states (singlet for Fe(II) and doublet for Fe(III)). Intermediate spin states were not considered. Oxidation states are qualitative and obtained by constraining total charge to correspond to the net charge on the respective ligands along with a positive (+2 or +3) charge for the iron center. Quantitative determination of the charges and occupation of subshells (i.e. $3d$ and $4s$) was obtained from the TeraChem interface with the Natural Bond Orbital (NBO) v6.0 package[71]. NBO calculates the natural atomic orbitals (NAOs) for each atom by computing the orthogonal eigenorbitals of the atomic blocks in the density matrix. After the set of NAOs is defined, NAO occupancy is obtained using natural population analysis (NPA)[72], which permits estimation of $3d$ and $4s$ subshell occupation. The NBO partial charge ($q$) on an atom is calculated by taking the difference between the atomic number ($Z$) and the total population ($N$) for the NAOs for each atom ($i$):

$$q_i = Z_i - N_i \ . \tag{6}$$

Several octahedral complex structures (ligands: CO, CN⁻, CNH, NCH, $NH_3$, $H_2O$) were generated from simplified molecular input line entry system (SMILES)[73] strings. Using OpenBabel[74], the SMILES strings were converted to structures that were starting points for TeraChem geometry optimizations. The larger octahedral complex structures (ligands: (phen)$_2$(NCS)$_2$, PEPXEP, HICPEQ, bpy, terpy), were obtained from the Cambridge Structural Database (CSD)[75]. PEPXEP denotes the CSD accession code for a compound with N$_6$C$_{26}$H$_{38}$ stoichiometry, while HICPEQ corresponds to a N$_8$C$_{18}$H$_{26}$ compound. The (phen)$_2$(NCS)$_2$ structure was previously identified as a good test case[51]. The other ligands were selected by using the CCDC ConQuest web-screening tool with a query limiting elements to Fe, C, N, H in an octahedral complex with symmetric Fe-N bonds, as was previously used for catalyst screening[76].

**IV. Dependence of spin-state ordering on functional choice**

We have considered a test set of representative Fe(II) and Fe(III) octahedral complexes (Fig. 1) for various exchange-correlation functionals. In all cases, the ground state spin is known experimentally or may be suggested from ligand field theory. Fe(II) and Fe(III) have nominally $3d^6$ and $3d^5$ electron configurations, giving rise to low-spin (LS) singlet or doublet spin multiplicity or high-spin (HS) quintet or sextet electronic states. The adiabatic electronic energy gap between HS and LS states is:

$$\Delta E^{\text{HS-LS}} = E_{\text{HS}}(\mathbf{R}_{\text{HS}}) - E_{\text{LS}}(\mathbf{R}_{\text{LS}}), \qquad (7)$$

where $E_{\text{HS}}(\mathbf{R}_{\text{HS}})$ is the electronic energy of the HS state at its geometry optimized coordinates and $E_{\text{LS}}(\mathbf{R}_{\text{LS}})$ is the equivalent for the LS state. The initial set of structures includes two carbon ligand sets (CO and CNH), three nitrogen ligand sets (NH$_3$, NCH, and (phen)$_2$(SCN)$_2$), and one oxygen ligand set (H$_2$O) (see structures in Fig. 1). One representative functional is chosen for each class: LDA (PZ81[37]), GGA (PBE[77]), GGA hybrid (B3LYP) and meta-GGA (M06-L[78]) to compare qualitative relative high-spin/low-spin energetics. Reliance on a single representative functional for each class is motivated by preliminary findings in comparing a wider array of functionals (see list in Sec. III). Pure density functionals (LDA or GGA) consistently predict low-spin ground states in nine of the ten cases (six are Fe(II) and four are Fe(III)) considered (Table 1), although only half of the ten cases are expected to be low spin.

Pure GGA preference for low-spin Fe(II)/Fe(III) complexes is consistent with earlier observations[31, 79, 80]. Including higher order dependence on the density as in a meta-GGA improves identification of some high-spin states: Fe(II)(NH$_3$)$_6$ and Fe(III)(NCH)$_6$ are predicted to be high spin with a meta-GGA, while they were predicted to be low-spin with a GGA. However, meta-GGA results are inconsistent: Fe(III)(NH3)$_6$ and Fe(II)(NCH)$_6$ have high-spin ground states[31, 81] but the meta-GGA predicts both to be low-spin. Identification of how the higher-order terms of the density may be systematically incorporated to improve predictions of magnetic ordering or spin states is of ongoing interest for future work because meta-GGAs have the potential to improve predictions in extended systems where explicit incorporation of Hartree-Fock exchange may be prohibitive. For the GGA hybrid class of functionals, correct qualitative identification of spin states is achieved in eight out of ten cases. However, in the case of (phen)$_2$(NCS)$_2$, a high-spin ground state is predicted despite experimental observation[51] of a low-spin ground state. While this test set is relatively small, it reinforces general observations that GGA hybrids tend to over-predict high-spin ground states, while pure density functionals predict low-spin ground states. This trend will be investigated on an expanded molecule test set in Sec. V.

Qualitative spin-state assignment is difficult in weak ligand cases where the quantitative gap falls below 5 kcal/mol due to basis set dependence or zero-point energy and vibrational entropy effects[31, 33] not considered here but covered in detail in the recent work by Mortensen and Kepp[33]. For the GGA, Fe(II)(NH$_3$)$_6$ is close to crossover to high-spin, which would improve agreement with experiment. Three of the meta-GGA predictions: Fe(II)(NCH)$_6$, Fe(III)(NH$_3$)$_6$, and Fe(II)(phen)$_2$(NCS)$_2$, are close to the spin crossover point to HS states, which would improve agreement with experiment in the first two cases but worsen agreement for the last case. We also note that these meta-GGA results may be more substantially sensitive to the functional form since M06-L, for instance, is highly parameterized. We thus compare against TPSS[82], a meta-GGA with fewer adjustable parameters. Comparing the TPSS and M06-L meta-GGAs, we find the two are qualitatively consistent, but TPSS has a stronger bias for high spin systems. This bias leads to improved qualitative agreement for two compounds (Fe(II)(NCH)$_6$ and Fe(III)(NH3)$_6$) as high-spin but also reduced qualitative agreement for two low-spin compounds that TPSS predicts to be high-spin (Fe(II)(phen)$_2$(NCS)$_2$ and Fe(III)(CO)$_6$).

**V. Dependence of spin-state ordering on HF exchange**

In order to broadly investigate the effect of HF exchange on spin-state ordering, we vary the amount of HF exchange included in a modified B3LYP (modB3LYP) functional. The DFT exchange for the modB3LYP functional is calculated using the following expression:

$$E_x^{modB3LYP} = a_{HF}E_x^{HF} + (1-a_{HF})E_x^{LDA} + 0.9(1-a_{HF})\left(E_x^{GGA} - E_x^{LDA}\right), \tag{8}$$

where $a_{HF}$ is the amount of HF exchange. For $a_{HF} \to 0$, the exchange is pure DFT-GGA (as in BLYP), while for $a_{HF} \to 1$, the exchange becomes pure HF. The factor 0.9 was introduced so that the ratio:

$$\frac{E_x^{GGA}}{E_x^{LDA}} = 9 \tag{9}$$

is equal to that of the original B3LYP functional (0.72 for $E_x^{GGA}$ and 0.08 for $E_x^{LDA}$) and constant for all $a_{HF}$. We apply the modB3LYP functional (with HF exchange = 0-50%) to select octahedral complexes from the initial test set (Sec. IV) as well as an expanded test set. Throughout, we also compare to a narrow range of 12.7-28.3%, which corresponds to a 3σ confidence interval on the normal distribution fit to the votes for standard hybrid exchange-correlation functionals in a popular density functional theory poll[2]. While the narrower range indicates the most common hybrid exchange ratios, the wider range permits connection to both pure GGA and high HF exchange functionals.

**A. Spin-state ordering dependence with Fe(III) complex test cases**

We first focus on the relative electronic energy between high-spin (HS) and low-spin (LS) electronic states ($\Delta E^{HS-LS}$) for four Fe(III) octahedral complexes (N ligands: NCH, NH$_3$, C ligands: CNH, CO) across the 0-50% HF exchange range (Fig. 2). Fe(III) complexes have a $d^5$ configuration that will lead to complete filling of all $d$ levels in the high-spin case or a paired, closed-shell doublet in the low-spin case. Linear behavior in spin-state energetics is observed over the complete range of HF exchange covered with modB3LYP exchange for Fe(III) complexes with both carbon and nitrogen ligands, extending and confirming earlier observations by Droghetti on octahedral Fe(II) complexes[31] over a range of about 15-40% HF exchange.

This linear energetic dependence is surprising because it suggests that any response that the density has to the modified HF potential is of the same magnitude in both the high-spin and low-spin state. That is, it is evident that simply mixing increasing fractions of HF exchange energy on a high-spin state will lower its energy linearly with respect to a low-spin state. However, if the density responds differently in the case of the low-spin state, e.g. through increased localization with respect to the high-spin state due to imbalances in self-interaction error, then the energetic dependence should contain higher order terms than a simple linear averaging. This linear result suggests that HF-derived localization of covalent, delocalized orbitals may not be suitable for understanding the effect of higher fractions of HF exchange. A comparison to self-interaction correction schemes[36, 37] and the delocalization-penalty +U approach[42] will likely be instructive in the future.

Since $\Delta E^{HS-LS}$ varies linearly with HF exchange, linear-regression fits are very good approximations to the partial derivative of the energy with respect to HF exchange ($a_{HF}$):

$$\text{slope} = \frac{\Delta \Delta E^{HS-LS}}{\Delta a_{HF}} \approx \frac{\partial \Delta E^{HS-LS}}{\partial a_{HF}} \quad (10)$$

where the correlation coefficients (i.e., $R^2$ values) for these fits are all 0.999. We introduce here the unit notation "HFX", where one unit of HFX corresponds to the range from 0 to 100% HF exchange. The identity of the directly bonded element dominates the value of $\frac{\partial \Delta E^{HS-LS}}{\partial a_{HF}}$, and nitrogen-containing ligands have nearly identical values: -75 $\frac{\text{kcal}}{\text{mol} \cdot \text{HFX}}$ for $Fe(III)(NH_3)_6$ and -77 $\frac{\text{kcal}}{\text{mol} \cdot \text{HFX}}$ for $Fe(III)(NCH)_6$. For carbon-containing ligands, the correspondence is also quite close: -110 $\frac{\text{kcal}}{\text{mol} \cdot \text{HFX}}$ for $Fe(III)(CNH)_6$ and -106 $\frac{\text{kcal}}{\text{mol} \cdot \text{HFX}}$ for $Fe(III)(CO)_6$. Over the $3\sigma$ confidence interval, the carbon ligand sets always prefer a low-spin ground state, but $\Delta E^{HS-LS}$ is reduced by 17 kcal/mol over this range, which is a significant change in predictions of quantitative energetics. For comparison, the shift in $\Delta E^{HS-LS}$ from BLYP (0%) to B3LYP (20%) is larger at around -21 kcal/mol, and the difference between a 20% and 25% HF exchange shifts the prediction by -5 kcal/mol. The ratio of Hartree-Fock exchange and the direct ligand dominate these trends, rather than the form of the DFT exchange or the associated correlation functional. When calculations are carried out with a modified PBE0 functional instead of modB3LYP, slopes are qualitatively unchanged, with an average difference of 6% in computed slopes and a maximum deviation of $-9 \frac{\text{kcal}}{\text{mol} \cdot \text{HFX}}$ for $Fe(III)(CO)_6$.

While the spin-state splitting derivatives are smaller for the nitrogen-containing ligands, the proximity of the curves to the HS-LS crossover makes the qualitative spin-state assignment more challenging. Furthermore, both $Fe(III)(NH_3)_6$ and $Fe(III)(NCH)_6$ are low-spin at the lower bound of the $3\sigma$ confidence interval ($a_{HF}$=0.127), while they are both high-spin at the upper bound of that same interval ($a_{HF}$=0.283). If the objective of a computational study is qualitative spin-state assignment, such an assignment would be highly sensitive to functional choice. Experimentally[81], $Fe(III)(NH3)_6$ is known to be high-spin, but HS-LS spin crossover occurs at 27.2% HF exchange, which is a higher fraction than is incorporated in B3LYP or PBE0. Despite challenges in qualitative assignment, quantitative changes in spin-state orderings are slightly lower: the shift in $\Delta E^{HS-LS}$ in the confidence interval is -12 kcal/mol, from BLYP to B3LYP it is -15 kcal/mol, and the difference between a 20% and 25% HF exchange is -4 kcal/mol.

Previous work in this area[33, 83-85] suggests that a ligand field theory (LFT) picture that focuses on ligand strength following the spectrochemical series[86] may provide some, albeit tenuous, guidance regarding observations in HS-LS splitting. The CO ligand is the strongest in the spectrochemical series and should maximize octahedral field splitting ($\Delta_o$) between the three low-energy $t_{2g}$ and two high-energy $e_g$ states, while the $NH_3$ ligand is considerably weaker and should have a smaller $\Delta_o$ value. Across the range of all HF exchange percentages, the LS state is relatively preferred for the strong CO with respect to the $NH_3$ ligand. However, for high HF exchange (40-50%), the HS state is the ground state for $Fe(III)(CO)_6$ and the relative penalty of $\Delta E^{HS-LS}(CO)$ vs. $\Delta E^{HS-LS}(NH_3)$ narrows. In a simplified LFT picture, increasing HF exchange is modulating the octahedral field splitting more dramatically for strong ligands (e.g. CO) than for weak ligands (e.g. $NH_3$). Thus, these trends suggest that too-high ratios of exact exchange in functionals will override established ligand field concepts.

**B. Spin-state ordering: comparison with Fe(II) complexes**

Qualitative trends in spin-state ordering with $a_{HF}$ (Fig. 3) previously observed for Fe(III) are preserved in Fe(II), but with slightly lower correlation coefficients ($R^2 = 0.995$-$0.997$). For Fe(II), carbon ligand systems have higher $\frac{\partial \Delta E^{HS\text{-}LS}}{\partial a_{HF}}$ values: -151 $\frac{\text{kcal}}{\text{mol} \cdot \text{HFX}}$ for Fe(II)(CNH)$_6$ and -155 $\frac{\text{kcal}}{\text{mol} \cdot \text{HFX}}$ for Fe(II)(CO)$_6$, and this higher slope appears to correlate with higher $\Delta E^{HS\text{-}LS}$ versus Fe(III) obtained at a GGA reference by 20 kcal/mol. Such a difference between Fe(II) and Fe(III) $\Delta E^{HS\text{-}LS}$ diverges from ligand field theory, since in LFT, Fe(II) does not populate any additional high-energy levels in the HS or LS state. The Fe(II)(NH$_3$)$_6$ and Fe(II)(NCH)$_6$ $\frac{\partial \Delta E^{HS\text{-}LS}}{\partial a_{HF}}$ values diverge slightly from their Fe(III) counterparts, reducing the slope to -63 $\frac{\text{kcal}}{\text{mol} \cdot \text{HFX}}$ in the former case and increasing to -86 $\frac{\text{kcal}}{\text{mol} \cdot \text{HFX}}$ in the latter. Nevertheless, trends are preserved: here, spin-crossover from LS to HS occurs near the lower bound of the 3σ confidence interval, while both ligands prefer high spin at the upper bound.

Despite high correlation coefficients, the fit to linear trend lines appears poorer in the case of Fe(II) compared to Fe(III). It is likely that extra degrees of freedom associated with the unpaired minority-spin 3$d$ electron in Fe(II) make energetic predictions more sensitive to HF exchange ratios. Quadratic fits of the data were thus also obtained, leading naturally to an improved fit. First derivatives of the second order polynomials give access to a range of instantaneous $\frac{\partial \Delta E^{HS\text{-}LS}}{\partial a_{HF}}$ values. By definition, the derivatives obtained at 25% HF exchange from either approach match exactly, but this single value is an underestimate for low $a_{HF}$ and an overestimate for high $a_{HF}$. The carbon ligand slope ranges are from -106 ($a_{HF}$=0.5) to -206 ($a_{HF}$=0.0) for CO and -114 ($a_{HF}$=0.5) to -183 ($a_{HF}$=0.0) for CNH. Nitrogen ligand ranges are slightly smaller: -48 to -81 for NH$_3$ and -50 to -125 for NCH. Such ranges are subject to the number of data points and nature of the higher order fit, but they provide a reference frame for evaluating the magnitude of variation of derivatives. Thus, although there is a 23 $\frac{\text{kcal}}{\text{mol} \cdot \text{HFX}}$ difference in $\frac{\partial \Delta E^{HS-LS}}{\partial a_{HF}}$ from linear regression for Fe(II)/N complexes, this difference is relatively small. As in the case of Fe(III), use of a modified PBE0 functional produces comparable slopes to those from modB3LYP. The average deviation in slopes between the two approaches is 5% with the largest deviation being -10 $\frac{kcal}{mol \cdot HFX}$ for Fe(II)(CO)$_6$. However, in light of the non-linearity analysis for Fe(II), it becomes clear that discrepancies between the two classes of tuned functionals are within the uncertainty of the slope assignment.

Direct comparison of Fe(II) and Fe(III) nitrogen ligand trends permits identification of the magnitude of differences between the oxidation states (Fig. 4). The $\Delta E^{HS-LS}$ splitting of Fe(II)(NCH)$_6$ and Fe(III)(NCH)$_6$ is nearly identical for data points in the 10-40% HF exchange range. Small differences in $\Delta E^{HS-LS}$ ~ 1-2 kcal/mol shift the prediction of the numerical slope -8 $\frac{kcal}{mol \cdot HFX}$ from Fe(III) to Fe(II). In contrast, $\Delta E^{HS-LS}$ shifts downward by nearly 20 kcal/mol from Fe(III)(NH$_3$)$_6$ to Fe(II)(NH$_3$)$_6$. The two NH$_3$ curves appear parallel for HF exchange below 20%, but HS stabilization rate declines at higher %HF exchange for Fe(II). Such observations suggest that $\frac{\partial \Delta E^{HS-LS}}{\partial a_{HF}}$ depends more strongly on direct ligand identity than on oxidation state. Therefore, prediction variability with exchange-correlation parameter changes may be determined on a small set and broadly applied to an array of compounds.

In order to investigate whether the correlations observed thus far hold for larger transition-metal complexes, we enlarge the data set for $\frac{\partial \Delta E^{HS-LS}}{\partial a_{HF}}$ evaluation (structures in Fig. 5). For the Fe(III) compounds, there is a narrow range of derivatives from about -70 to -80 $\frac{kcal}{mol \cdot HFX}$ for the six nitrogen ligand sets and -105 to -110 $\frac{kcal}{mol \cdot HFX}$ for the three carbon ligand sets. The agreement for direct-ligand-based spin-state splitting dependence on exact exchange is less strong for Fe(II) compounds: four nitrogen complexes (NH$_3$, bpy, terpy, and PEPXEP) have values around -110 to -120 $\frac{kcal}{mol \cdot HFX}$ while NH$_3$ and NCH are -63 and -86 $\frac{\text{kcal}}{\text{mol} \cdot \text{HFX}}$, respectively. For the Fe(II) carbon ligand sets, CO and CNH slopes are around -155 $\frac{\text{kcal}}{\text{mol} \cdot \text{HFX}}$ while CN is slightly shallower at -130 $\frac{\text{kcal}}{\text{mol} \cdot \text{HFX}}$. For almost all data points, the Fe(II) $a_{\text{HF}}$ gradients are larger in magnitude than the Fe(III) $a_{\text{HF}}$ gradients.

**C. Trends in charge localization measures**

Self-interaction-error induced delocalization is one rationale for why pure density functionals fail to correctly predict relative spin-state energetics in transition metal complexes. Localized 3d electrons are expected to be particularly sensitive to SIE, and low-spin states permit greater delocalization through higher occupancy of bonding orbitals than high-spin states do. In order to quantify the extent of charge localization on the transition metal center, we compute formal charges with NBO natural population analysis. The difference in charge obtained between the HS and LS states is then a predictor of the relative charge-localization between the states:

$$\Delta q^{\text{HS-LS}} = q_{\text{Fe}}^{\text{HS}} - q_{\text{Fe}}^{\text{LS}}, \tag{11}$$

where the charge on Fe for all cases considered is positive (i.e. less than the atomic number of iron). A positive $\Delta q^{\text{HS-LS}}$ corresponds to a net loss of electrons on the iron center from LS to HS states. For all 18 cases considered (both Fe(II) and Fe(III)), increasing the % of HF exchange increases the formal positive charge on both HS and LS states, which indicates an absence of charge localization on the Fe center through the inclusion of HF exchange. Instead, this effective delocalization of charge from the metal to neighboring ligands opposes the view on how HF exchange corrects SIE on transition metal valence states. A second metric is the dependence of the charge on percentage of HF exchange:

$$\text{slope} = \frac{\Delta \Delta q^{\text{HS-LS}}}{\Delta a_{\text{HF}}} \approx \frac{\partial \Delta q^{\text{HS-LS}}}{\partial a_{\text{HF}}}, \tag{12}$$

where this derivative is obtained from linear regression of charges with $a_{\text{HF}}$. Positive values for the charge difference derivative, $\frac{\partial \Delta q^{\text{HS-LS}}}{\partial a_{\text{HF}}}$, indicate that the HS state loses electrons to surrounding ligands faster than the LS state, and a negative value corresponds to the reverse.

First, correlations between spin-state HF exchange dependence and the shift in charge from the LS to HS state ($\Delta q^{HS-LS}$) are compared for Fe(III) compounds (Fig. 6). Charge reduction from LS to HS is greater in the case of carbon (~1.2 e$^-$ loss) than nitrogen (~0.6-0.8 e$^-$ loss) ligands. Derivatives of $\Delta E^{HS-LS}$ with respect to HF exchange demonstrate a strong correlation ($R^2$=0.94) to the charge shift ($\Delta q^{HS-LS}$) evaluated at 20% HF exchange. That is, the more charge lost from the LS to the HS state on the iron center, the greater the stabilization of the HS state as the fraction of exact exchange is increased. The resulting linear-scaling relation from the least squares fit is:

$$\frac{\partial \Delta E^{HS-LS}}{\partial a_{HF}} \approx -73.5 \Delta q^{HS-LS}_{@20\%} - 19.6. \tag{13}$$

Extrapolation beyond the surveyed compounds may not be on firm footing, and collection of results on additional ligand sets could alter this linear-scaling relation. Nevertheless, this relationship suggests that if the relative charge increases on the HS state, then the dependence of the splitting will be reduced. In particular, a stationary point (i.e. $\frac{\partial \Delta E^{HS-LS}}{\partial a_{HF}} = 0$) may be extrapolated to occur at a $\Delta q^{HS-LS}$ of -0.27. That is, if the high-spin Fe(III) state accumulates charge with respect to the low-spin state, then increasing HF exchange from a pure GGA should not affect the HS-LS splitting. It is likely difficult to isolate such systems because covalent interactions are stronger in LS iron states than HS states, leading to a higher net charge. Nevertheless, this observation provides a search direction for identification of Fe(III) complexes that have relatively inert qualitative and quantitative spin-state orderings with respect to HF exchange. While it may be preferable to continue to develop electronic structure methods that are increasingly accurate, narrowing the focus of materials screening and discovery efforts to exchange-correlation-inert compounds is a potential direction to circumvent many of the uncertainties faced in applications of practical DFT. This narrowing still leaves a wide chemical space open that is inclusive of many potentially catalytically active iron complexes.

It is worthwhile to see whether these correlations hold for Fe(II) compounds (Fig. 7),

recalling that for Fe(II) the dependence of spin-state splitting on HF exchange exhibited more variability and non-linear behavior. A comparable correlation is found for the Fe(II) compounds, although with a reduced correlation coefficient ($R^2 = 0.67$). Three Fe(II) complexes (NH$_3$, NCH, and CN) lie above the best-fit line, and a shallower dependence on charge would be obtained if those three points were excluded. Using all nine data points, a linear-scaling relationship is obtained:

$$\frac{\partial \Delta E^{HS-LS}}{\partial a_{HF}} \approx -67.7 \Delta q^{HS-LS}_{@20\%} - 47.0. \tag{14}$$

Despite the reduced quality of the fit for Fe(II) complexes with respect to Fe(III), the linear-scaling relationship is consistent. The second derivative of spin-state splittings with respect to both HF exchange and charge (-73.5 vs. -67.7 $\frac{\text{kcal/mol}}{\text{HFX} \cdot -e^-}$) appears invariant to oxidation state of the transition metal and direct ligand identity. One difference for Fe(II) is the prediction of the stationary point for spin-state splitting with respect to HF exchange at a higher charge increase on the HS state of $\Delta q^{HS-LS} = -0.7\, e^-$.

Now considering all 18 complexes, $\frac{\partial \Delta E^{HS-LS}}{\partial a_{HF}}$ is plotted against the partial derivative of the charge with respect to HF exchange, $\frac{\partial \Delta q^{HS-LS}}{\partial a_{HF}}$, in Fig. 8. Both positive and negative values of $\frac{\partial \Delta q^{HS-LS}}{\partial a_{HF}}$ (from -1.2 to 0.2 $\frac{-e^-}{\text{HFX}}$) are observed. Positive charge shift derivatives indicate that HS states lose charge faster than LS states with increasing $a_{HF}$ and are associated with Fe(III)-nitrogen complexes. The Fe(III)/N complexes also have the flattest dependence of spin-state splittings on HF exchange, suggesting that if increasing HF exchange causes depletion of charge in the HS state with respect to the LS state, the HF-exchange derived stabilization of the HS state is reduced.

Fe(II)-carbon complexes exhibit opposite behavior, with charge depleting more slowly from the HS state with increasing $a_{HF}$ and larger dependence of $\Delta E^{HS-LS}$ on HF exchange. The Fe(III)/C and Fe(II)/N complexes are intermediate in their values of charge dependence on HF exchange. For the best fit line, data from all oxidation state/ligand combinations is included, excluding four outlying points: Fe(II)(NH$_3$)$_6$, Fe(II)(NCH)$_6$, Fe(II)(terpy)$_2$, and Fe(II)(CN)$_6$. Exclusion of these four compounds leads to a strong correlation ($R^2$=0.93) for the remaining 14 data points with a linear-scaling relationship:

$$\frac{\partial \Delta E^{HS-LS}}{\partial a_{HF}} \approx 106 \frac{\partial \Delta q^{HS-LS}}{\partial a_{HF}} - 84. \tag{15}$$

This relationship provides quantitative support for the observation that $a_{HF}$-inert complexes are most probable for cases where charge accumulates on the HS state with increasing HF exchange. A stationary point for $\Delta E^{HS-LS}$ may be extrapolated to $\frac{\partial \Delta q^{HS-LS}}{\partial a_{HF}} = 0.8 \frac{-e^-}{HFX}$. This quantitative result suggests that a GGA will give the same spin-state ordering as a hybrid if increasing HF exchange causes increased localization of charge on the HS state with respect to the LS state. This result is somewhat surprising because when SIE is invoked for suggesting a hybrid functional over a GGA for 3$d$ electrons, it is assumed that HF exchange should always enhance localization in a high-spin state. However, few compounds in the data set have positive derivatives of charge shift with respect to HF exchange. Identification of putative compounds that extend the set surveyed thus far would validate this prediction.

We now provide justification for exclusion of the four outliers from the linear regression in Fig. 8. Fe(II)(NH$_3$)$_6$, Fe(II)(NCH)$_6$, and Fe(II)(CN)$_6$ were previously observed to have lower than expected spin-state splitting dependence on HF exchange (see Fig. 7). The linear fit to obtain $\frac{\partial \Delta q^{HS-LS}}{\partial a_{HF}}$ for Fe(II)(NH$_3$)$_6$ and Fe(II)(CN)$_6$ is poorer with respect to other data points because first derivatives increase abruptly around 20-30% HF exchange. Earlier analysis also identified that depending on the evaluation point (e.g. at lower HF exchange), values of $\frac{\partial \Delta E^{HS-LS}}{\partial a_{HF}}$ for NCH and NH$_3$ ligands could be evaluated as significantly higher than values obtained at 25% HF exchange. For Fe(II)(CN)$_6$, a net relative increase in 3$d$ occupations in the HS state over the LS state for increasing HF exchange ($\frac{\partial \Delta 3d^{HS\text{-}LS}}{\partial a_{HF}}$) is double that observed for other compounds (1.2 e⁻/HFX vs. 0.6 e- e⁻/HFX for other compounds). Generally, both HS and LS states lose 3$d$ occupancy with increasing $a_{HF}$ and the LS state has higher overall 3$d$ occupancy than the HS state. For Fe(II)(CN)$_6$, the LS state loses electrons much more rapidly than the HS state, suggesting that HF exchange has a much stronger effect on the LS state than the HS state. For Fe(II)(terpy)$_2$, this data point is omitted to avoid asymmetric exclusion of outlying data points, which would skew the trend line. One further justification is that the Fe(II)(terpy)$_2$ complex was challenging to converge and there is some increased variability in the fit for $\frac{\partial \Delta q^{HS\text{-}LS}}{\partial a_{HF}}$ as a result. In future work, finer-grained data for partial derivatives and exclusion of data points that are subject to higher internal uncertainties could lead to clarification of the trend line versus designation of outliers.

**D. Corroborating geometric and energetic relationships**

In all studies, we observe consistent increased variability in both energetic and charge descriptions of Fe(II) octahedral complexes over Fe(III) complexes. There also appears to be a stronger dependence of energetic properties on HF exchange for carbon ligands with respect to nitrogen ligands. Structural correlations ($\Delta R^{HS\text{-}LS}$ in Table 2 and the partial derivative in Table 3) support the charge and energetic trends. As expected, bond elongation occurs for octahedral complexes in the HS state, but the range of $\Delta R^{HS\text{-}LS}$ at 20% HF exchange over the Fe(II)/carbon complexes is largest both in magnitude (avg=0.38 Å) and range (max-min=0.08 Å). This $\Delta R^{HS\text{-}LS}$ shift is smallest for Fe(III)/nitrogen complexes (avg=0.16 Å), and the range is also reduced. The partial derivative of bond length shifts with respect to HF exchange is negative in all cases with comparable averages across all compounds of about -0.15-0.09 $\frac{\text{Å}}{\text{HFX}}$. The Fe(II)/carbon octahedral complexes again have the highest degree of variability. Negative derivatives in the case of all complexes indicate that the inclusion of increasing percentages of HF exchange makes the formal bond order between HS and LS states more equivalent. In Fe(II) complexes, this behavior corresponds to bond elongation in the LS state and unchanging or slight decrease for HS state bond lengths, suggesting that HF exchange is reducing covalent hybridization in LS state bonding orbitals. The majority of Fe(III) complexes instead exhibit negative derivatives due to decreases in bond length for the high-spin complex, indicating stronger bonding, and unchanging bond lengths in the low-spin complex.

### E. Quantitative vs. qualitative spin-state ordering

In earlier spin-state dependence observations (see Figs. 5 and 6), qualitative spin-state assignment and quantitative spin-state dependence on HF exchange appeared decoupled. That is, if HS and LS states were nearly degenerate at 20% HF exchange, the variation of those spin-state splittings with respect to HF exchange was reduced compared to cases where the HS-LS splitting was high. We now consider the strength of this correlation on all 18 octahedral complexes in this study (Fig. 9). Over the data set, HS complexes at 20% HF exchange have the smallest dependence of splittings on HF exchange. Such a result suggests that stable HS complexes should be more exact-exchange inert. When the 18 octahedral complexes are grouped by oxidation state, relatively good correlations are observed in the fit for all nine Fe(III) complexes ($R^2$=0.80) and nine Fe(II) complexes ($R^2$=0.72). In both cases, a stationary point for HF exchange dependence of spin-state splitting is predicted at a $\Delta E^{HS-LS}$ around -65 to -73 kcal/mol. Such compounds would be both quantitatively and qualitatively identified as high-spin and would likely correspond to weak field ligand interactions. A recent study[33] of eight density functionals provides further validation that stronger ligands are more sensitive to differences in functional parameters including, but not limited to, HF exchange mixing[33, 83-85]. Weak ligands that can test the exchange inert ranges should be discoverable, as $\Delta E^{HS-LS}$ in isolated $Fe^{2+}$ and $Fe^{3+}$ gas phase ions is -86 and -134 kcal/mol, respectively[87]. Another target for extrapolation from linear scaling relations is that the energy should change no more than 2 kcal/mol (i.e., ± 1 kcal/mol chemical accuracy with respect to the midpoint of the normal distribution) over the 3$\sigma$ confidence interval (12.7-28.3% HF exchange) in the functional. Using this measure, such narrow uncertainties would occur for octahedral complexes with relative HS spin state energetics at around -53 to -64 kcal/mol.

### VI. Conclusions

In this work, we have quantified ranges of property prediction for transition metal complexes based on exchange-correlation functional choice over a range of the most commonly used functional properties (i.e. GGA hybrids with a typical HF exchange range (3σ) of 12.7-28.3%). Despite the proliferation of a "zoo" of functionals, we observed qualitative agreement for various functionals within a functional class when comparing spin-state ordering across ten representative Fe(II) and Fe(III) octahedral complexes with various ligands. We used a fixed GGA/LDA ratio modified B3LYP functional to study the dependence of properties across the 3σ interval and beyond (0-50%). With increasing HF exchange, we observed strong high-spin stabilization over low-spin complexes, as much as 1-2 kcal/mol per 1% HF exchange. High HF exchange (> 30%) even overrode qualitative ligand field theory arguments by stabilizing high-spin, strong-ligand $Fe(CO)_6$ complexes. While HS-LS energetics depend strongly on HF exchange, the strength of variation was linear in nature and broadly applicable across oxidation state, and varied most with respect to the element of the direct ligand. These observations suggest that HF-exchange-dependence may be straightforwardly determined across broad classes of many materials by a handful of calculations on representative molecules.

We then identified the extent to which tuning HF exchange changes the underlying charge density. We made the surprising observation that partial charge decreases on iron as HF exchange is increased, corresponding to $3d$ electron delocalization to ligands. This runs counter to typical explanations of how HF exchange can approximately correct self-interaction errors. Further, we identified a general correlation between HS-LS splitting with HF exchange and HS-LS charge differences and first derivatives, suggesting that when the charges are more balanced between HS and LS states, the energy dependence on HF exchange should be reduced. This work echoes prior observations of the lack of a one-size-fits-all percentage of HF exchange for transition-metal complexes. Nevertheless, we have identified relative uncertainties in spin-state ordering that correlate well with broad metal and ligand identities, and we have identified weak field ligands and balanced HS-LS partial charges to be good objectives for materials to consider in catalyst and materials design because molecules with such properties will have reduced sensitivity to exchange-correlation functional choice.  Future work will focus on refining descriptors for functional tuning and charge matching against wavefunction theory.

**Corresponding Author**

*E-mail: hjkulik@mit.edu

**Acknowledgments**

H.J.K. holds a Career Award at the Scientific Interface from the Burroughs Wellcome Fund. The authors (H.J.K. and E.I.I.) were supported by an MIT Energy Initiative seed grant.

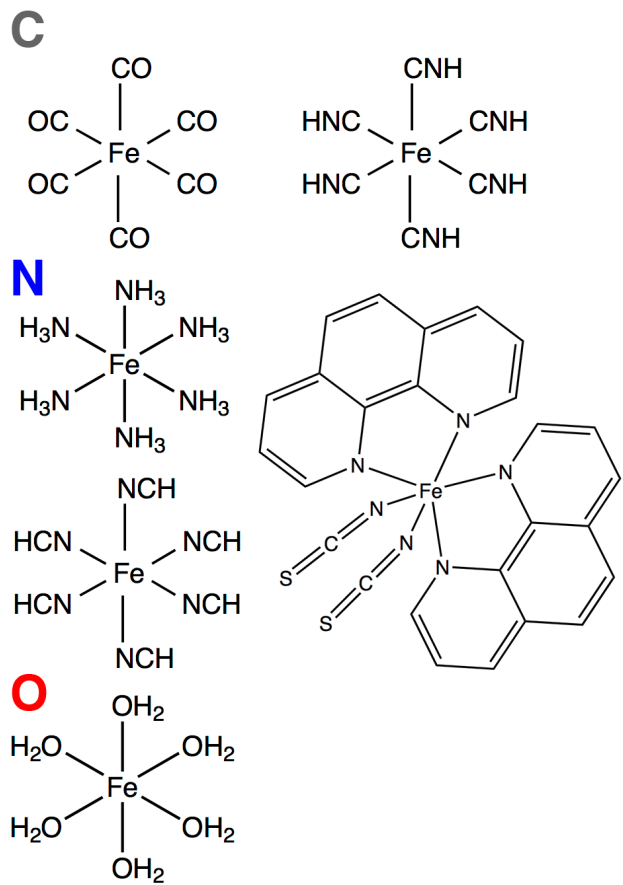

**Figure 1.** Structures of octahedral iron complexes classified by direct ligand identity: carbon (top), nitrogen (middle), or oxygen (bottom).

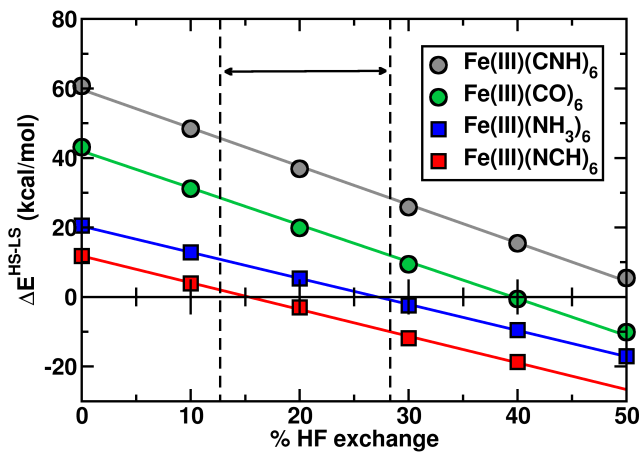

**Figure 2.** Relative high spin (HS)-low spin (LS) energy ($\Delta E^{HS-LS}$) in kcal/mol of four Fe(III) octahedral complexes (two nitrogen ligands: NCH and $NH_3$ and two carbon ligands: CNH and CO) with $3\sigma$ confidence interval from normal distribution poll data on hybrid exchange functionals, as indicated with black dashed lines and black arrow.

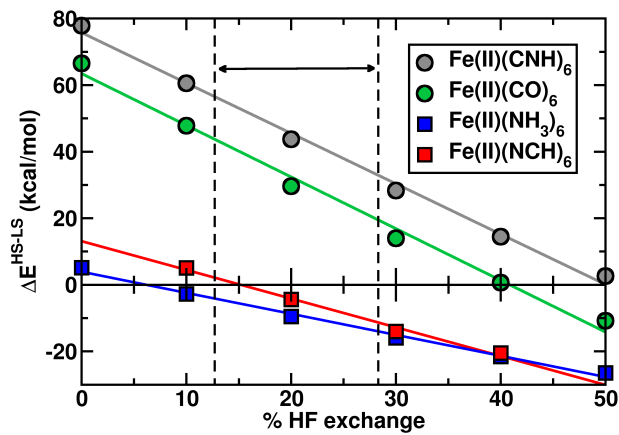

**Figure 3.** Relative high spin (HS)-low spin (LS) energy ($\Delta E^{HS-LS}$) in kcal/mol of four Fe(II) octahedral complexes (two nitrogen ligands: NCH and $NH_3$ and two carbon ligands: CNH and CO) with $3\sigma$ confidence interval from normal distribution poll data on hybrid exchange functionals, as indicated with black dashed lines and black arrow.

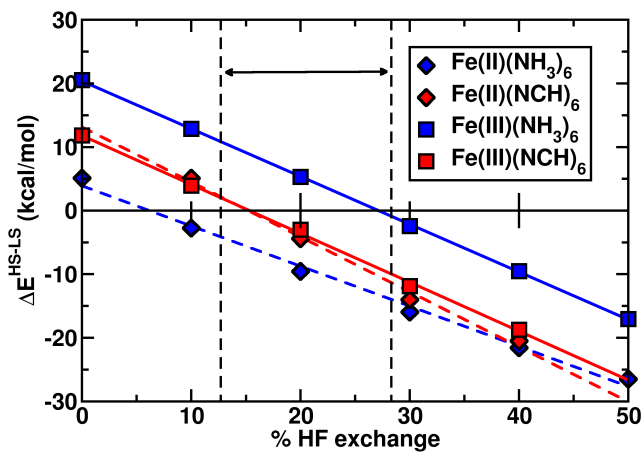

**Figure 4.** Relative high spin (HS)-low spin (LS) energy ($\Delta E^{HS-LS}$) in kcal/mol of octahedral complexes with nitrogen ligands (NCH, indicated in red and $NH_3$, indicated in blue) for both Fe(II) (diamond symbols) and Fe(III) (square symbols) with $3\sigma$ confidence interval from normal distribution poll data on hybrid exchange functionals, as indicated with black dashed lines and black arrow. The trend lines for Fe(II) are dashed lines, while the trend lines for Fe(III) are solid lines.

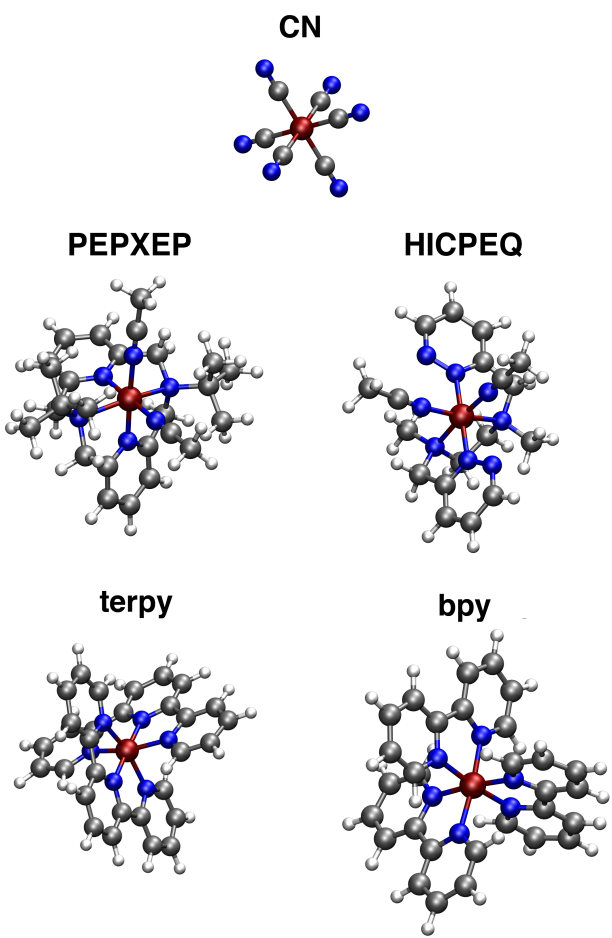

**Figure 5.** Ball-and-stick models of nitrogen- and carbon-ligand structures for octahedral iron complexes with compound labels used throughout text. The atoms are color-coded with nitrogen in blue, carbon in gray, hydrogen in white, and iron in maroon.

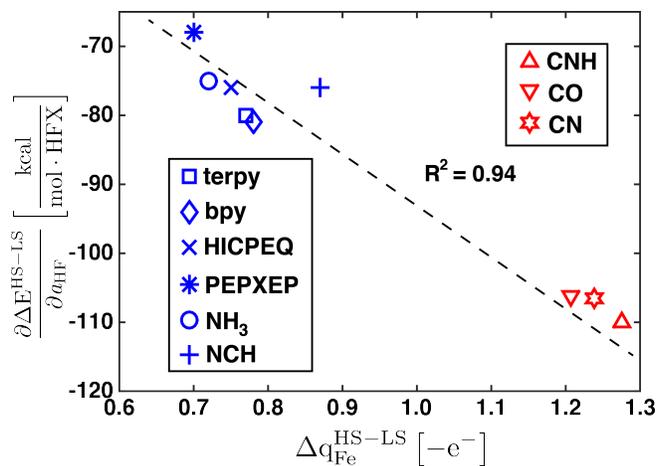

**Figure 6.** Plot of the derivative of spin-state splitting with Hartree-Fock exchange ($a_{HF}$) against the difference in Fe(III) NBO charges between high-spin (HS) and low-spin (LS) states. Results are shown for six nitrogen containing ligands (blue symbols) and three carbon containing ligands (red symbols), as indicated with legend. A linear regression fit and associated $R^2$ value is also shown on the plot.

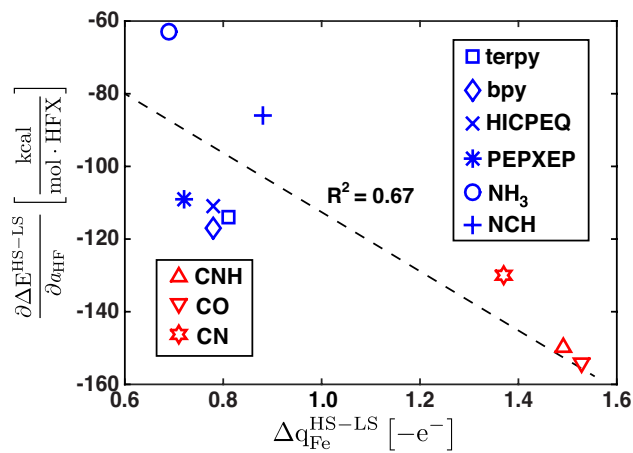

**Figure 7.** Plot of the derivative of spin-state splitting with Hartree-Fock exchange ($a_{HF}$) against the difference in Fe(II) NBO charges between high-spin (HS) and low-spin (LS) states. Results are shown for six nitrogen containing ligands (blue symbols) and three carbon containing ligands (red symbols), as indicated with legend. A linear regression fit and associated $R^2$ value is also shown on the plot.

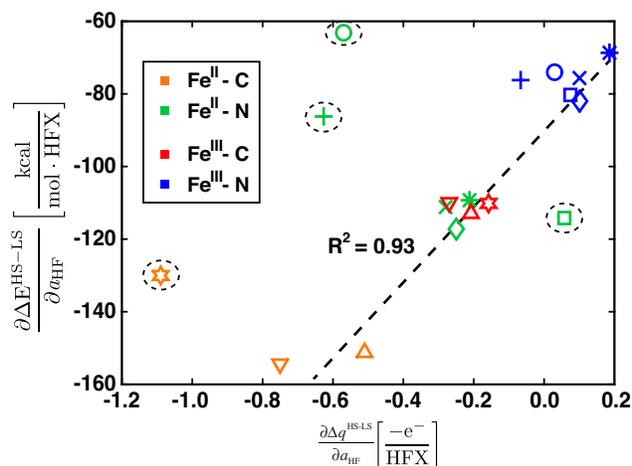

**Figure 8.** Plot of the derivative of spin-state splitting with Hartree-Fock exchange ($a_{HF}$) against the high-spin-low-spin NBO charge difference derivative with $a_{HF}$. Results are shown for Fe(II) and Fe(III) with six nitrogen containing ligands (green symbols for Fe(II), blue for Fe(III)) and three carbon containing ligands (orange symbols for Fe(II), red for Fe(III)). Symbol shapes follow the legend in Figs. 9 and 10. A linear least-squares regression fit is indicated with a dashed line, with four outlying data points excluded from the fit as indicated by dashed outer circles.

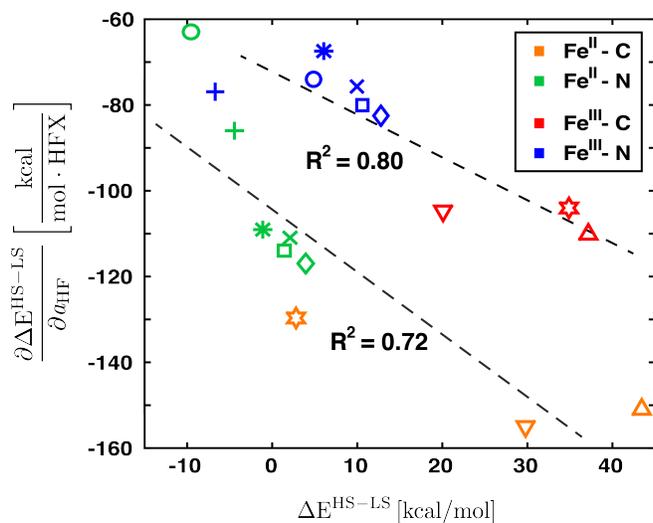

**Figure 9.** Plot of the derivative of spin-state splitting with Hartree-Fock exchange ($a_{HF}$) against spin-state splitting at 20% HF exchange. Results are shown for Fe(II) and Fe(III) with six nitrogen containing ligands (green symbols for Fe(II), blue for Fe(III)) and three carbon containing ligands (orange symbols for Fe(II), red for Fe(III)). Symbol shapes follow the legend in Figs. 9 and 10. Two best-fit lines are obtained for Fe(II) and Fe(III) compounds separately, as indicated by dashed lines.

**Table 1.** Ground states of octahedral Fe(II) and Fe(III) complexes with specified ligand sets for LDA, GGA, GGA hybrid, and meta-GGA classes of exchange-correlation functionals. Reference (Ref.) data is from experiment (indicated in bold), otherwise approximations from ligand-field theory are provided. Incorrect predictions of the ground state spin for a functional are indicated by italics. The experimental data is from those collected in Ref. [31], except for Fe(II)(H$_2$O)$_6$ (Ref. [88]), Fe(II/III)(NH$_3$)$_6$ (Ref. [81]), and Fe(II)(phen)$_2$(NCS)$_2$ (Ref. [51]).

| Ligand | Ref. | LDA | GGA | hybrid | meta-GGA |
|---|---|---|---|---|---|
| | | Fe(II) | | | |
| (CO)$_6$ | **LS** | LS | LS | LS | LS |
| (CNH)$_6$ | **LS** | LS | LS | LS | LS |
| (NCH)$_6$ | **HS** | *LS* | *LS* | HS | *LS* |
| (NH$_3$)$_6$ | **HS** | *LS* | *LS* | HS | HS |
| (phen)$_2$(NCS)$_2$ | **LS** | LS | LS | *HS* | LS |
| (H$_2$O)$_6$ | **HS** | HS | HS | HS | HS |
| | | Fe(III) | | | |
| (CO)$_6$ | LS | LS | LS | LS | LS |
| (CNH)$_6$ | LS | LS | LS | LS | LS |
| (NCH)$_6$ | HS | *LS* | *LS* | HS | HS |
| (NH$_3$)$_6$ | **HS** | *LS* | *LS* | *LS* | *LS* |

**Table 2.** Difference in bond length at 20% Hartree-Fock exchange between high-spin (HS) and low-spin (LS) complexes. The average (avg), minimum (min), and maximum (max) differences are reported for each class of metal-ligand complexes.

| M-L class | ΔR(HS-LS) (Å) | | |
|---|---|---|---|
| | avg | min | max |
| Fe(II)-C | 0.38 | 0.34 | 0.42 |
| Fe(II)-N | 0.21 | 0.20 | 0.23 |
| Fe(III)-C | 0.26 | 0.25 | 0.28 |
| Fe(III)-N | 0.16 | 0.15 | 0.17 |

**Table 3.** Derivative of bond length differences between high-spin (HS) and low-spin (LS) complexes with respect to Hartree-Fock exchange. The average (avg), minimum (min), and maximum (max) differences are reported for each class of metal-ligand complexes.

| M-L class | $\frac{\partial \Delta R}{\partial HFX}$ (HS-LS) (Å) | | |
| --- | --- | --- | --- |
| | avg | min | max |
| Fe(II)-C  | -0.15 | -0.28 | -0.05 |
| Fe(II)-N  | -0.12 | -0.17 | -0.10 |
| Fe(III)-C | -0.14 | -0.18 | -0.11 |
| Fe(III)-N | -0.09 | -0.11 | -0.08 |

**References**

1.	K. Burke, J. Chem. Phys. **136**, 150901 (2012).

2.	M. Swart, F. M. Bickelhaupt, and M. Duran, Popularity poll of DFT functionals, http://www.marcelswart.eu/dft-poll (2014).

3.	Y. Zhao, and D. G. Truhlar, Chem. Phys. Lett. **502**, 1 (2011).

4.	N. Mardirossian, and M. Head-Gordon, Phys. Chem. Chem. Phys. **16**, 9904 (2014).

5.	C. Adamo, and V. Barone, J. Chem. Phys. **110**, 6158 (1999).

6.	M. Ernzerhof, and G. E. Scuseria, J. Chem. Phys. **110**, 5029 (1999).

7.	L. A. Curtiss, K. Raghavachari, P. C. Redfern, and J. A. Pople, J. Chem. Phys. **106**, 1063 (1997).

8.	A. J. Cohen, P. Mori-Sánchez, and W. Yang, Science **321**, 792 (2008).

9.	C. J. Cramer, and D. G. Truhlar, Phys. Chem. Chem. Phys. **11**, 10757 (2009).

10.	M. C. Gutzwiller, Phys. Rev. **134**, A923 (1964).

11.	J. N. Harvey, Annu. Rep. Prog. Chem., Sect. C: Phys. Chem. **102**, 203 (2006).

12.	S. Lutfalla, V. Shapovalov, and A. T. Bell, J. Chem. Theory Comput. **7**, 2218 (2011).

13.	K. P. Jensen, and U. Ryde, J. Biol. Chem. **279**, 14561 (2004).

14.	A. Jain, G. Hautier, S. P. Ong, C. J. Moore, C. C. Fischer, K. A. Persson, and G. Ceder, Phys. Rev. B **84** (2011).

15.	J. N. Harvey, R. Poli, and K. M. Smith, Coord. Chem. Rev. **238**, 347 (2003).

16.	P. Gütlich, and A. Hauser, Coord. Chem. Rev. **97**, 1 (1990).

17.	H.-J. Lin, D. Siretanu, D. A. Dickie, D. Subedi, J. J. Scepaniak, D. Mitcov, R. Clérac, and J. M. Smith, J. Am. Chem. Soc. **136**, 13326 (2014).

18. P. Gütlich, and H. A. Goodwin, *Spin crossover in transition metal compounds I* (Springer Science & Business Media, 2004).

19. J. A. Real, E. Andrés, M. C. Muñoz, M. Julve, T. Granier, A. Bousseksou, and F. Varret, Science **268**, 265 (1995).

20. L. Bogani, and W. Wernsdorfer, Nat. Mater. **7**, 179 (2008).

21. S. Sanvito, Chem. Soc. Rev. **40**, 3336 (2011).

22. D. Schröder, S. Shaik, and H. Schwarz, Acc. Chem. Res. **33**, 139 (2000).

23. R. Poli, and J. N. Harvey, Chem. Soc. Rev. **32**, 1 (2003).

24. Y. H. Kwon, B. K. Mai, Y.-M. Lee, S. N. Dhuri, D. Mandal, K.-B. Cho, Y. Kim, S. Shaik, and W. Nam, J. Phys. Chem. Lett. **6**, 1472 (2015).

25. K. Yoshizawa, Y. Shiota, and T. Yamabe, Chem. Eur. J. **3**, 1160 (1997).

26. Y. Shiota, and K. Yoshizawa, J. Am. Chem. Soc. **122**, 12317 (2000).

27. M. Filatov, and S. Shaik, J. Phys. Chem. A **102**, 3835 (1998).

28. H. J. Kulik, and N. Marzari, J. Chem. Phys. **129**, 134314 (2008).

29. H. J. Kulik, M. Cococcioni, D. A. Scherlis, and N. Marzari, Phys. Rev. Lett. **97**, 103001 (2006).

30. H. J. Kulik, and N. Marzari, in *Fuel Cell Science: Theory, Fundamentals, and Bio-Catalysis*, edited by J. Norskov, and A. Wiezcowski (Wiley monograph, 2010), pp. 433.

31. A. Droghetti, D. Alfè, and S. Sanvito, J. Chem. Phys. **137**, 124303 (2012).

32. G. Ganzenmüller, N. Berkaïne, A. Fouqueau, M. E. Casida, and M. Reiher, J. Chem. Phys. **122**, 234321 (2005).

33. S. R. Mortensen, and K. P. Kepp, J. Phys. Chem. A (2015).

34. S. Zein, S. A. Borshch, P. Fleurat-Lessard, M. E. Casida, and H. Chermette, J. Chem. Phys. **126**, 014105 (2007).

35. M. Swart, A. R. Groenhof, A. W. Ehlers, and K. Lammertsma, J. Phys. Chem. A **108**, 5479 (2004).

36. M. R. Pederson, A. Ruzsinszky, and J. P. Perdew, J. Chem. Phys. **140**, 121103 (2014).

37. J. P. Perdew, and A. Zunger, Phys. Rev. B **23**, 5048 (1981).

38. K. P. Jensen, and J. Cirera, J. Phys. Chem. A **113**, 10033 (2009).

39. T. F. Hughes, and R. A. Friesner, J. Chem. Theory Comput. **7**, 19 (2010).

40. F. Neese, JBIC, J. Biol. Inorg. Chem. **11**, 702 (2006).

41. H. Paulsen, V. Schünemann, and J. A. Wolny, Eur. J. Inorg. Chem. **2013**, 628 (2013).

42. H. J. Kulik, J. Chem. Phys. (invited perspective, submitted 2015).

43. A. Ghosh, B. J. Persson, and P. Taylor, JBIC Journal of Biological Inorganic Chemistry **8**, 507 (2003).

44. F. Aquilante, P.-Å. Malmqvist, T. B. Pedersen, A. Ghosh, and B. O. Roos, J. Chem. Theory Comput. **4**, 694 (2008).

45. A. Ghosh, E. Gonzalez, E. Tangen, and B. O. Roos, J. Phys. Chem. A **112**, 12792 (2008).

46. K. Pierloot, and S. Vancoillie, J. Chem. Phys. **128**, 034104 (2008).

47. M. Radoń, E. Broclawik, and K. Pierloot, The Journal of Physical Chemistry B **114**, 1518 (2010).

48. S. Vancoillie, H. Zhao, M. Radoń, and K. Pierloot, J. Chem. Theory Comput. **6**, 576 (2010).

49. L. M. Lawson Daku, F. Aquilante, T. W. Robinson, and A. Hauser, J. Chem. Theory Comput. **8**, 4216 (2012).

50. A. Vargas, I. Krivokapic, A. Hauser, and L. M. Lawson Daku, Phys. Chem. Chem. Phys. **15**, 3752 (2013).

51. M. Reiher, Inorg. Chem. **41**, 6928 (2002).

52. M. Reiher, O. Salomon, and B. A. Hess, Theor. Chem. Acc. **107**, 48 (2001).

53. A. Fouqueau, S. Mer, M. E. Casida, L. M. Lawson Daku, A. Hauser, T. Mineva, and F. Neese, J. Chem. Phys. **120**, 9473 (2004).

54. A. Fouqueau, M. E. Casida, L. M. L. Daku, A. Hauser, and F. Neese, J. Chem. Phys. **122**, 044110 (2005).

55. A. D. Becke, Phys. Rev. A **38**, 3098 (1988).

56. C. Lee, W. Yang, and R. G. Parr, Phys. Rev. B **37**, 785 (1988).

57. A. D. Becke, J. Chem. Phys. **98**, 5648 (1993).

58. P. J. Stephens, F. J. Devlin, C. F. Chabalowski, and M. J. Frisch, J. Phys. Chem. **98**, 11623 (1994).

59. J. Toulouse, F. Colonna, and A. Savin, Phys. Rev. A **70**, 062505 (2004).

60. T. Stein, J. Autschbach, N. Govind, L. Kronik, and R. Baer, J. Phys. Chem. Lett. **3**, 3740 (2012).

61. R. Baer, E. Livshits, and U. Salzner, Annu. Rev. Phys. Chem. **61**, 85 (2010).

62. J. H. Skone, M. Govoni, and G. Galli, Phys. Rev. B **89**, 195112 (2014).

63. H. J. Kulik, J. Chem. Phys. **142**, 240901 (2015).

64. T. Weymuth, and M. Reiher, Int. J. Quantum Chem. **115**, 90 (2015).

65. Petachem., http://www.petachem.com

66. S. Vosko, L. Wilk, and M. Nusair, Can. J. Phys. **58**, 1200 (1980).

67. J. P. Perdew, Phys. Rev. B **33**, 8822 (1986).

68.	J. P. Perdew, and Y. Wang, Phys. Rev. B **45**, 13244 (1992).

69.	A. D. Becke, J. Chem. Phys. **107**, 8554 (1997).

70.	J. Kästner, J. M. Carr, T. W. Keal, W. Thiel, A. Wander, and P. Sherwood, J. Phys. Chem. A **113**, 11856 (2009).

71.	NBO6.0., E. D. Glendening, J, K. Badenhoop, A. E. Reed, J. E. Carpenter, J. A. Bohmann, C. M. Morales, C. R. Landis, and F. Weinhold, Theoretical Chemistry Institute, University of Wisconsin, Madison (2013).

72.	A. E. Reed, R. B. Weinstock, and F. Weinhold, J. Chem. Phys. **83**, 735 (1985).

73.	D. Weininger, J. Chem. Inf. Comput. Sci. **28**, 31 (1988).

74.	N. O'Boyle, M. Banck, C. James, C. Morley, T. Vandermeersch, and G. Hutchison, J. Cheminform. **3**, 33 (2011).

75.	F. H. Allen, Acta Crystallogr. Sect. B **58**, 380 (2002).

76.	H. J. Kulik, S. E. Wong, S. E. Baker, C. A. Valdez, J. Satcher, R. D. Aines, and F. C. Lightstone, Acta Crystallogr. Sect. C **70**, 123 (2014).

77.	J. P. Perdew, K. Burke, and M. Ernzerhof, Phys. Rev. Lett. **77**, 3865 (1996).

78.	Y. Zhao, and D. G. Truhlar, J. Chem. Phys. **125**, 194101 (2006).

79.	A. Sorkin, M. A. Iron, and D. G. Truhlar, J. Chem. Theory Comput. **4**, 307 (2008).

80.	F. Furche, and J. P. Perdew, J. Chem. Phys. **124**, 044103 (2006).

81.	D. N. Bowman, and E. Jakubikova, Inorg. Chem. **51**, 6011 (2012).

82.	J. Tao, J. P. Perdew, V. N. Staroverov, and G. E. Scuseria, Phys. Rev. Lett. **91**, 146401 (2003).

83.	C. Anthon, and C. E. Schäffer, Coord. Chem. Rev. **226**, 17 (2002).

84.	C. Anthon, J. Bendix, and C. E. Schäffer, Inorg. Chem. **42**, 4088 (2003).

85. J. Moens, G. Roos, P. Jaque, F. De Proft, and P. Geerlings, Chem. Eur. J. **13**, 9331 (2007).

86. W. Moffitt, and C. Ballhausen, Annu. Rev. Phys. Chem. **7**, 107 (1956).

87. A. Kramida, Y. Ralchenko, J. Reader, and the, NIST ASD Team, (2014).

88. P. Gütlich, Y. Garcia, and H. A. Goodwin, Chem. Soc. Rev. **29**, 419 (2000).